\documentclass[letterpaper]{article} 
\usepackage{aaai24}  
\usepackage{times}  
\usepackage{helvet}  
\usepackage{courier}  
\usepackage[hyphens]{url}  
\usepackage{graphicx} 
\usepackage{natbib}  
\usepackage{caption} 
\usepackage{algorithm}
\usepackage{algorithmic}
\usepackage{amsmath}
\usepackage{booktabs}
\usepackage[dvipsnames]{xcolor}
\usepackage{colortbl}

\usepackage{newfloat}
\usepackage{listings}
\floatstyle{ruled}
\newfloat{listing}{tb}{lst}{}
\floatname{listing}{Listing}
\usepackage{xspace}

\newcommand{\algo}{ArguSense\xspace}
\newcommand{\duration}{21\xspace}  
\newcommand{\DS}{DIS\xspace}
\newcommand{\DSlong}{Deliberation Intensity Score\xspace}

\newcommand{\Dargument}{$D_{Arg}$\xspace}
\newcommand{\Dcluster}{$D_{Cluster}$\xspace}

\newcommand{\showComments}{\input{001-Notes-Comments.tex}} 

\newcommand{\deli}{deliberation\xspace}

\newcommand{\ArguPerc}{27} 

\newcommand{\WordsPerPost}{50\xspace} 
\newcommand{\PostLengthPerc}{50} 

\newcommand{\ThreadsAll}{$T_{all}$\xspace} 
 \newcommand{\ThreadsGMO}{$T_{GMO}$\xspace} 
\newcommand{\ThreadsLong}{$T_{Long}$\xspace} 
\newcommand{\ThreadsXLong}{$T_{XLong}$\xspace} 

\newcommand{\miii}[1]{{\textcolor{purple}{\bf MF:}} 
{\textcolor{Maroon}{\bf #1}}}
\newcommand{\miok}[1]{{\textcolor{purple}{\bf MF:}} 
{\textcolor{ForestGreen}{\bf #1}}}

\newcommand{\kevin}[1]{{\textcolor{green}{\bf KE:}} 
{\textcolor{red}{\bf #1}}}
\newcommand{\kevok}[1]{{\textcolor{green}{\bf KE:}} 
{\textcolor{Plum}{\bf #1}}}

\newcommand{\arman}[1]{{\textcolor{green}{\bf AI:}} 
{\textcolor{red}{\bf #1}}}
\newcommand{\armok}[1]{{\textcolor{green}{\bf AI:}} 
{\textcolor{CadetBlue}{\bf #1}}}

 \newcommand{\eatreminders}{
  \renewcommand{\miii}[1]{}
  \renewcommand{\miok}[1]{##1}
   \renewcommand{\kevin}[1]{}
   \renewcommand{\kevok}[1]{##1}
   \renewcommand{\arman}[1]{}
   \renewcommand{\armok}[1]{##1}
   \renewcommand{\armok}[1]{##1}
    \renewcommand{\showComments}{}
   }
 
\eatreminders

\usepackage{xcolor}

\title{\algo: Argument-Centric Analysis of Online Discourse}
\author{
    Arman Irani,
    Michalis Faloutsos,
    Kevin Esterling
}
\affiliations{
    University of California, Riverside \\
    
    airan002@ucr.edu, michalis@cs.ucr.edu, kevin.esterling@ucr.edu

}

\begin{document}

\maketitle

\begin{abstract}

How can we model arguments and their dynamics in online forum discussions?
The meteoric rise of online forums presents researchers across different disciplines with an unprecedented opportunity:
we have access to texts containing discourse between groups of users generated in 
a voluntary and organic fashion.
Most prior work so far has focused on classifying individual \textit{monological} comments as either argumentative or not argumentative. However, 
few efforts quantify and describe the \textit{dialogical processes} between users found in online forum discourse: the structure and content of interpersonal argumentation. Modeling dialogical discourse requires the ability to identify the presence of arguments, group them into clusters, and summarize the content and nature of clusters of arguments within a discussion thread in the forum. 
In this work, we develop \algo, a comprehensive and systematic \miok{framework} for understanding arguments and debate in online forums. Our \miok{framework consists} of methods for, among other things: (a) detecting argument topics in an unsupervised manner; (b) describing the structure of arguments within threads with powerful visualizations; and (c) quantifying
 the content and diversity of threads using argument similarity and clustering algorithms. 
We showcase our approach by analyzing the discussions
of four communities on the Reddit platform 
over a span of \duration months.
Specifically, we analyze the structure and content of threads related to GMOs in forums related to agriculture or farming to demonstrate the value of our \miok{framework}.  

\end{abstract}

\section{Introduction}

How can we measure the structure and content of argumentation in online discourse? This is the question at the heart of our work.

Online platforms and social media serve as a primary venue for discourse over public policies and other topics in modern society. Social media sites are said to place users in echo chambers of ideologically like-minded content \citep{Flaxman2016-it}, but threads of posts in online discussion forums often vary in the variety of perspectives and the intensity and dynamics of argumentation \citep{Friess2015-wu}. Given the importance of idea sharing in online platforms in contemporary society \citep{Sunstein2017-st}, it is important to develop measures of the nature and diversity of arguments and perspectives which users actually observe. 

While one can measure many aspects of discourse in online forums, in this paper we focus specifically on the presence and process of \textit{arguments}, a concept of interest that spans the fields of linguistics, communication, philosophy and political science \citep{Chesnevar2006-lq}. 
Formally, an argument is defined as the conjunction of at least one \textit{claim} with at least one \textit{premise} to justify the claim \citep{Yanal1991-wt}. Understanding the arguments posited in online forums help us to understand not only the opinions users hold on topics, but also the reasoning for why they hold those opinions \citep{Mercier2012-ui}.

\begin{figure}[t] 
\centering
\begin{minipage}{0.79\linewidth}
  \includegraphics[scale=0.10]{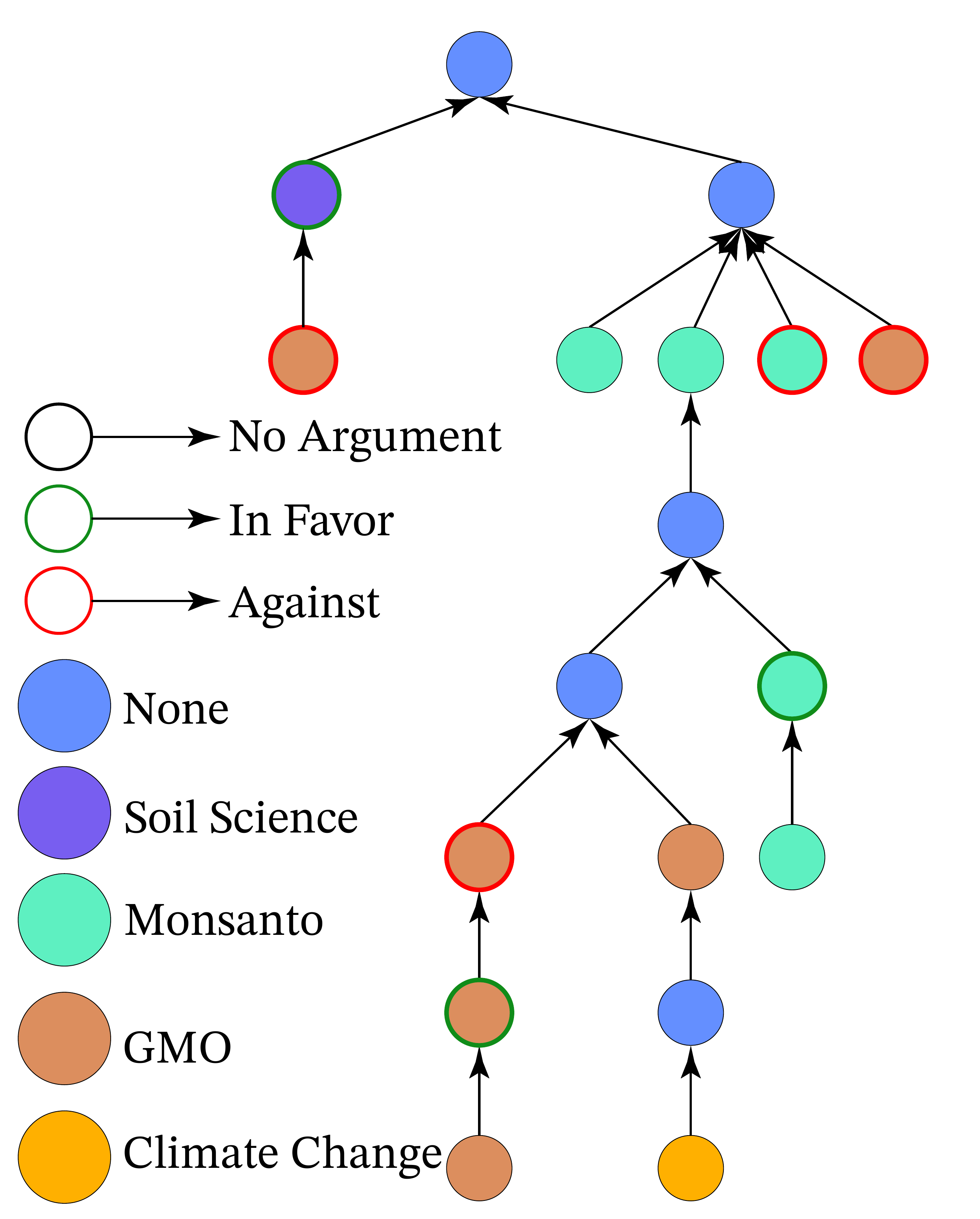}
  \centering
\end{minipage}
\caption{A visual representation of the \deli of a real Reddit thread around the GMO debate: 18 posts, 13 aspect-based posts, 7 argumentative posts, 4 different aspects, depth of 7, and a fan out of 7.
The color of a node corresponds to an aspect and the color of the border to the stance towards that aspect.
}

\label{figs:thread_tree}
\end{figure}

{\bf Problem Definition:} The problem we address in this paper is as follows: how can we measure the properties of online argumentation effectively around a topic of interest? Our input is an online forum, \miii{and a topic, captured by a set of keywords.} The desired output is the dominant arguments and the interplay of these arguments. 
The ultimate goal is to understand the dynamics among the different arguments to discern the function each argument plays in \deli. For example, one argument could be brought up consistently and repeatedly by many different users, while another argument could be an ``ending'' argument that is left without a counter argument in the deliberation \citep{Boschi2021-km}. 

Our work focuses on online discourse on Reddit, which provides certain challenges for mining arguments since the writing is short and informal \citep{Dutta2020-hr}. Unlike Twitter-like broadcasting, discussion forums consist of \textit{threads}, which naturally maps to group discussion. These discussions may be non-linear, as they might be centered around a topic set by the first post, but then users are able to splinter off into 'sub-threads' where discussion around other topics may occur.

{\bf Previous work:} In the section on Related Work below, we describe our relationship to the vast literature on argument mining found in computational linguistics and computer and social science. In short, while there is a variety of approaches for classifying and extracting arguments at the individual or \textit{monological} level of argumentation \citep[for recent reviews, see][]{Lawrence2020-na, Schaefer2021-yt}, there is a relative dearth of methods for assessing the nature and quality of the \textit{dialogical process} that occurs between users, and particularly so for the special case of online forums \citep[but see][]{Chakrabarty2020-eq}. Below we describe how we leverage monological methods of unsupervised argument classification to build measures of the dialogical processes in Reddit threads, and we show how our work is a complement to related efforts to quantify and describe the dialogical processes in forums.

{\bf Contribution:}
We develop \algo, a systematic \miok{framework} for quantifying
the extent and nature of argumentation in online discussion forums around a concept of interest. 
In a nutshell,  \algo consumes the unstructured data of a forum, and outputs meaningful insights as to the intensity and breadth of arguments in the \deli of the threads. 
Our  \miok{framework} consists of the following synergistic modules, each of which comes with non-trivial technical challenges:
(a) we identify the topical aspects around which arguments are formed for a concept;
(b) we identify the relevant threads to the concept and aspects of interest;
(c) we identify the argument(s) (or lack thereof) in each post, which includes the main aspect and the stance;
(d) we represent the discussion with an annotated semantically-rich tree
and develop metrics to characterize the \deli;
(e) we provide 
an intelligent content-aware method to cluster similar arguments;
(f) we propose a measure of deliberative intensity of threads, which is a weighted sum of the diversity of arguments and the amount of argumentation within a thread.
In each module, we adapt and customize state-of-the-art methods. 
However, the key novelty is the seamless integration 
of these tools
and the overarching conceptual framework of measuring the nature and extent of argumentation.
\miii{We had the two previous phrases already in the paper: let's just report it again in our review report}

We demonstrate the capabilities of \algo by applying it to 200,000+ posts across \duration months of 
agriculture-focused sub-Reddit forums.

We use the term \textbf{topic} to define a controversial debate, and we consider here the debate around Genetically Modified Organisms (GMOs). We use the term \textbf{aspect} to refer to concepts that represent arguments associated with the topic of interest. For example, the aspects are Monsanto, Climate Change, Soil Science appear often in a GMO discussion. 
\miok{It is important to note that our framework can be applied to any topic, especially if that topic can easily be captured in a small set of keywords.}
\miii{The previous phrase was already here, I just changed it a bit.}Note that our approach can leverage keyword expansion methods, and identification of related topics helps to automate its deployment by minimizing the need for manual input. Some of these capabilities are discussed later.

We highlight a few of the main observations from our study below.

{\bf a. Argumentation happens:  \ArguPerc\% of posts contain arguments.} We find that across all the posts of the GMO threads \ArguPerc\% of them contain an argument. At the same time, posts without an argument may still contribute to the discussion, as they can convey: (a) agreement or disagreement, or (b) provide information, e.g., point to a related article. 

\begin{figure}[t] 
\centering
\begin{minipage}{0.99\linewidth}
  \includegraphics[width=\linewidth]{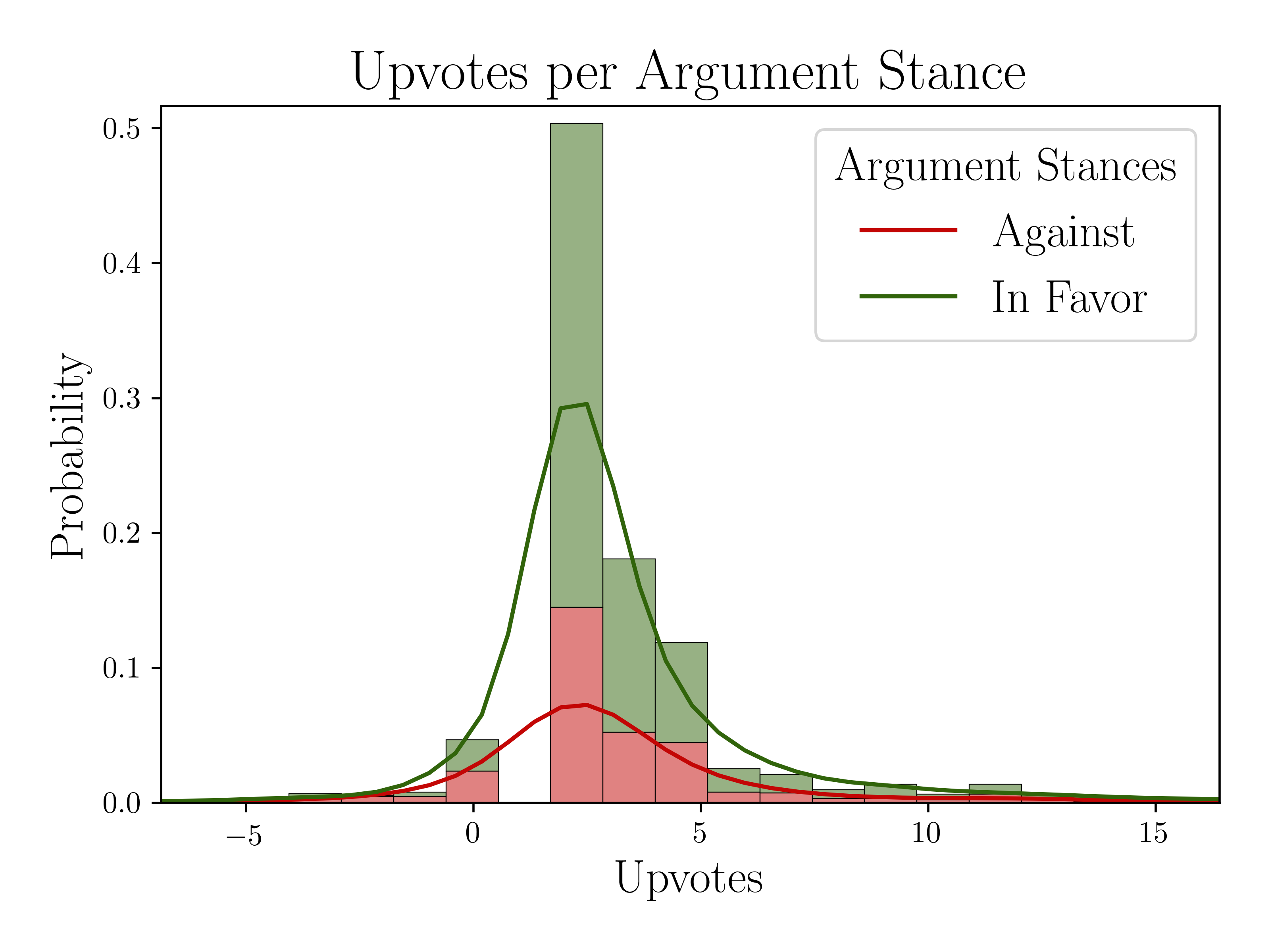}
  \centering
\end{minipage}
\caption{Positive stance posts get more upvotes: Upvote probability distribution based upon the stance of the argument. We do not count the default first upvote that every post on Reddit receives.}  
\label{figs:upvotes}
\end{figure}

\textbf{b. Identifying echo chamber behavior.}
We observe two phenomena that suggest the presence and reinforcement of echo chambers: (1) \textit{A post is more likely to reply to a post with the same stance and aspect as them if the argument is against.} Arguments against an aspect have a 19\% higher chance of being responded to by another argument against an aspect. This indicates echo chambers with against stance arguments occurring. Conversely, in favor aspect arguments have an equal likelihood of both stances responding, encouraging cross stance interactions.
(2) \textit{Arguments In Favor of an aspect are more likely to be upvoted.} Users are considerably more likely to upvote posts with an in favor argument than an argument against, as seen in Figure \ref{figs:upvotes}. This dialogical behavior promotes the visibility of in favor arguments over against arguments.

{\bf  c. Users put effort in their arguments: \PostLengthPerc\% of argumentative posts have more than \WordsPerPost words.} We observe that argumentative posts represent substantial effort. This point seems to suggest that online forums could have more informative discussions, especially if we compare it with posts of other social media, like Twitter. 

\textbf{d. Deliberation intensity does not increase with the size of the thread.} Observing the diversity of arguments and argument clusters within the conversation on GMOs allow us to quantify the intensity of deliberation. 
We propose a \DSlong metric, \DS,  to capture the dynamic nature of a deliberation. The metric is on a scale from 0 to 1, with 1 being intensely deliberative. We find 85\% of posts have a \DSlong of 0.2 or less, while only 5\% of threads have a \DSlong of more than 0.5. 
Contrary to intuition, we find that longer threads do not 
exhibit higher \DSlong. 
This indicates our \DS metric captures aspects of the  intensity of the debate in a way that goes beyond counting the number of posts. We explain how the DIS metric is calculated in the Methodology section.

\section{Data and Definitions}
This work will explore the extent and nature of argumentation in online forums. Therefore, we must define what we mean by ``arguments'' and we consider how the environment specific to Reddit influences how arguments are expressed.

{\bf What is an Argument?}
In the field of informal logic, an \textit{argument} is a statement that contains a claim supported by at least one premise \citep{Isenmann1997-wx}.  For example, an argument might be, (a) pesticides harm the environment, (b) GMOs reduce pesticide use, and so (c) GMOs are beneficial to the environment. In this argument, a) and b) are {\bf premises}, while (c) is a {\bf claim}. The three statements taken together can be classified as an argument. While the claim (c) may appear to be argumentative, without the related premises, (c) is only an opinion.

\setlength{\tabcolsep}{3pt}
\begin{table}[h]
\centering
\begin{tabular}{|l|l|l|l|}
\hline
\textbf{Forum} &  \# Threads &  \# Posts &  \# Users
\\ \hline
 r/vegetablegardening & 19,249 & 115,914 & 16,995+ \\ \hline
 r/farming & 12,441 & 69,683 & 11,198+ \\ \hline
 r/agriculture & 4,525 & 5,876 & 3,035+  \\ \hline
 r/horticulture & 2,792 & 14,521 & 3,822+  \\ \hline
\end{tabular}

\caption{Reddit dataset description.}
\label{tab:forums}
\end{table}

The study of arguments, and their proliferation and diversity, is important in public opinion research \citep{Mercier2012-ui}. In our modern media landscape, online forums such as Reddit have a significant impact on public discourse and opinion \citep{Sunstein2017-st}. Understanding the structures of communication in these mediums help policymakers and researchers keep up with rapidly evolving methods of persuasion and debate, and mining forums for the variety of arguments across topics on a policy issue can be informative to academics, government officials, media outlets, and the general public about the diversity of opinions and reasoning within communities of interest. 

We should be clear that in this paper, we do not consider the \textit{validity} of arguments \citep[see][]{Yanal1991-wt}. An argument is \textit{internally invalid} if (and only if) the claims do not follow from the premises, irrespective of whether the premises are themselves true, such as if statement (c) were rewritten to say GMOs are harmful to the environment. Importantly, one does not need to agree with an argument to acknowledge its internal validity. An argument is \textit{externally invalid} if (but not only if) the premises are false, for example if premise (b) was factually inaccurate. Generally speaking, invalid arguments are a subset of misinformation, the classification of which is outside the scope of this paper. However, because our methods identify and describe arguments in unstructured text, our results could help human labelers sift through text to identify arguments that can be assessed for their validity \citep{Alnemr2020-ii}.

{\bf Online forums: Reddit.}
We consider Reddit to be a valuable source of information. It is a popular online deliberation site, where millions of people come together to share information and opinions.
The platform consists of sub-forums or {\bf subreddits}, which we can think of as a community. Each subreddit consists of {\bf threads},
which function as discussions: a thread can be started by post of a user. 
Given the site's open nature, any user can participate in any thread
within any subreddit.
Therefore, there is typically a diverse range of opinions, beliefs and sub-interests represented. Naturally, there will be conflict and debate over specific topics and generally these topics focus on current events of ongoing public interest. A user can express their support or disagreement towards a post by ``upvoting'' or ``downvoting,'' with the numbers of up and down votes visible for each post. These votes can be taken as a community-led stamp of credibility and agreement \citep[but see][]{Ciampaglia2018-bw}.

{\bf The GMO debate.}
Here we focus on the debate around Genetically Modified Organisms or GMOs. The topics and keywords have been 
provided by researchers at our university who study GMOs. We explain later how we identify the related aspects by combining external input and automatically discovering some aspects.
We list all the topics and aspects in the beginning of our Case Study section.

\begin{figure}[t] 
\centering
\begin{minipage}{0.99\linewidth}
  \includegraphics[width=\linewidth]{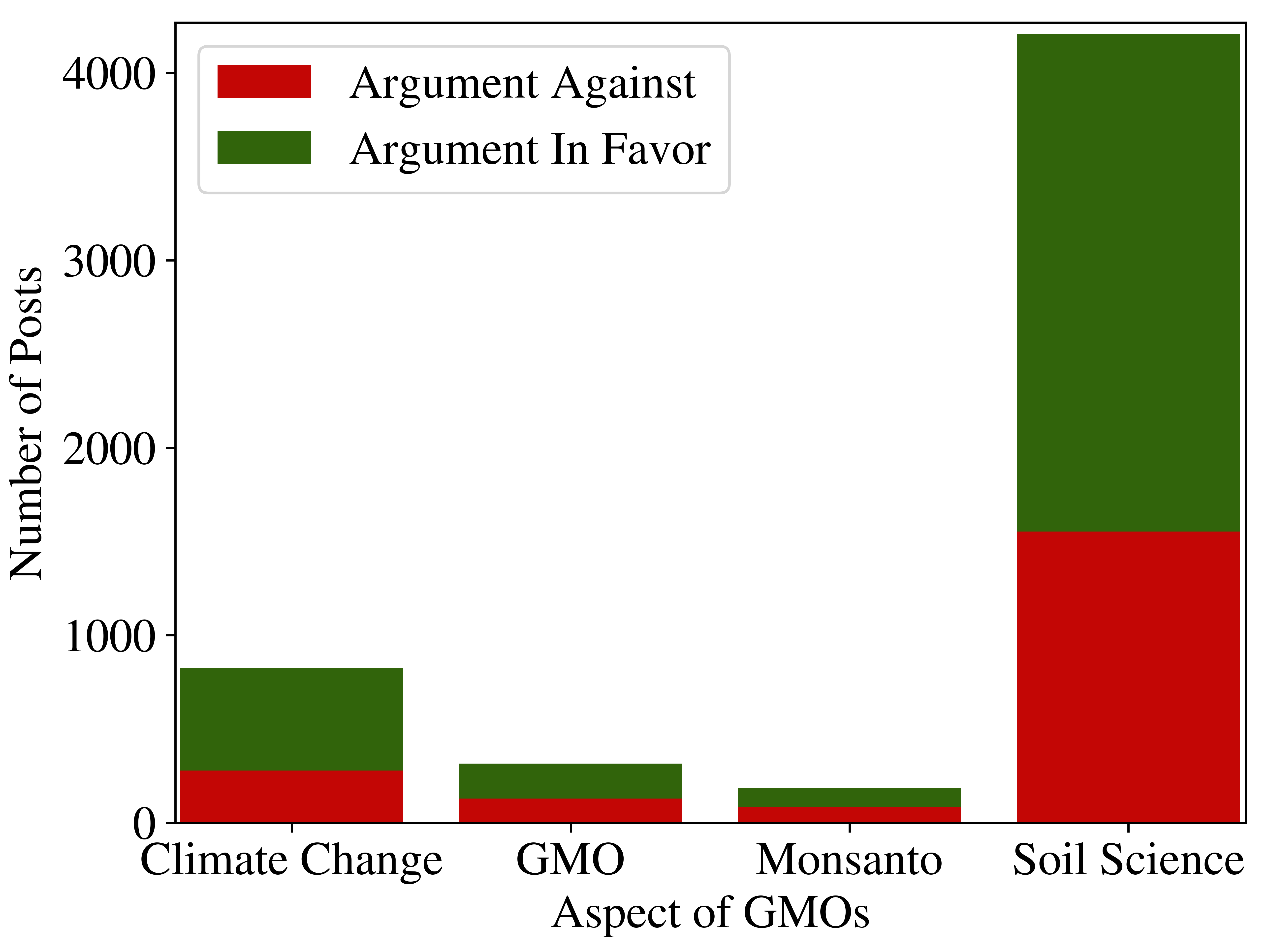}
  \centering
\end{minipage}
\caption{Breakdown of Argumentative Posts in our \ThreadsGMO dataset. Soil Science dominates the argumentative space, but there remains a relatively equal distribution of argumentative stance for all GMO aspects.}  
\label{figs:histogramposts}
\end{figure}

{\bf Our datasets.} For our analysis, we use data collected from four subreddit's, r/agriculture, r/horticulture, r/vegetablegardening, and r/farming. These four communities each represent a sector of the industry. 
r/agriculture consists of discussion regarding crops and livestock for consumption and the industry around it. r/horticulture is a science based community that discusses methodologies behind crop cultivation. r/VegetableGardening contains hobbyists who engage in conversations regarding small-scale and personal gardening. r/Farming has a user base of large-scale farmers who specialize in monoculture farming (barley, wheat, corn, soy), and the techniques and equipment used to do so. Combined we have a total of 205,000+ posts and 35,000+ unique users participating in these subreddits between January 2019 and September 2020. 
For ease of reference, we can define the following sets of threads:

\ThreadsAll: All the threads that we collected from our forums.

\ThreadsGMO: All the threads that we find to be relevant to the GMO debate as we explain later.

\ThreadsLong: We select the threads in \ThreadsGMO that have more than 5 posts, which leads to 586 threads.

\ThreadsXLong: We select the threads in \ThreadsGMO  that have more than 10 posts, which leads to 74 threads.

Subsetting our data for longer threads creates the opportunity to examine more extensive discussions, which we want to do when we do argument clustering and cluster summaries, or assess the \DSlong.

\section{Methodology}
\begin{figure*}[t] 
\centering
\begin{minipage}{0.99\linewidth}
  \includegraphics[scale=0.14]{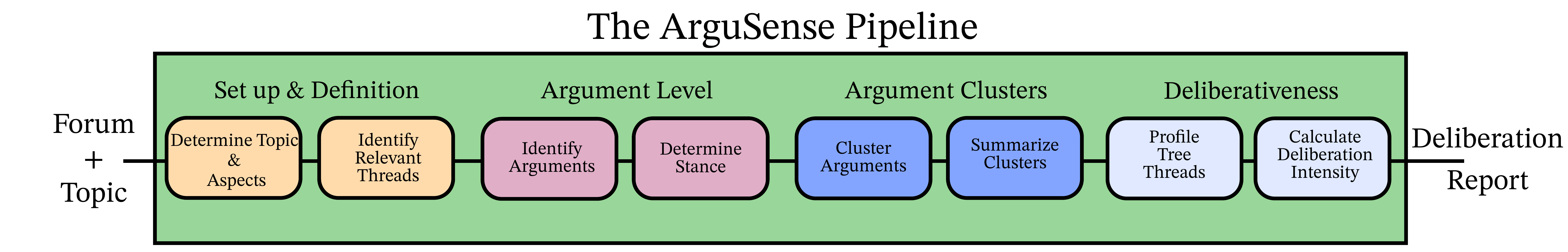}
  \centering
\end{minipage}
\caption{A visual representation of our methodology: the major phases and key capabilities.
}
\label{figs:methodology}
\end{figure*}

Our methodology consists of many different algorithms and capabilities. For ease of presentation,
we discuss them at three inter-connected levels:
(a) arguments,
(b)  groups of arguments,
and  (c) threads. We provide a visual representation of our methodology in Figure~\ref{figs:methodology}.

    \subsection{A. Our Argument Level Methods}

Here we propose measures to mine individual arguments within threads, for now setting aside the dialogical process between users. Our tasks are to identify the topic of the threads and to classify posts as either argumentative or not argumentative. 

\subsubsection{Aspect detection.}

As we mentioned in the introduction, we use the term \textit{aspect} to refer
to concepts that are commonly used in arguments for the topic of interest. In fact, these aspects can thought of as ``anchors" of  different arguments. E.g. the concept ``cancer" in a GMO discussion could anchor arguments that claim that GMOs cause or do not cause cancer.
The reason for this is two fold. We want to use the set of aspects in:
(a) identifying relevant threads,
and 
(b) determining if a post contains an argument.
We discuss both these functions below.

The aspects to a topic can be provided in three ways:
(a) given by a domain expert,
(b) identified algorithmically, and (c) both of the above.
There are several methods that can identify aspects related to a topic.
A recent method,  Flair \citep{akbik2019flair}, is considered the state-of-the-art NLP framework developed with PyTorch and a pretrained 
{\bf Named Entity Recognition (NER)} model, available under the MIT License, which is shown to perform very well. We use this method in our study.
For initial thread posts, we consider both the title and content of the post. 
Using the pretrained NER model, we classify each word in the post as either a person, location, organization, or other entity.

{\bf Identifying topic-relevant threads. }
At this stage, we assume that we have the topic and the aspects of interest. As an optimization step, we focus only on the threads that are relevant to the discussion to improve performance and avoid ``noise" from irrelevant threads. We therefore select threads that match our topic of interest as follows.
We identify all the likely-relevant posts: posts that contain words of interest, namely keywords of the topic or the aspects.
We then select the threads that have at least one likely-relevant post.

{\bf Argument  and stance identification.}
Having identified threads of interest, we classify topic-related posts as making an argument or not making an argument.
As monological argument mining is a relatively established area, we 
adopt and use  a technique that relies on contextualized BERT-based word embeddings \citep{ukpclassification} available under the Apache 2.0 License. 
For a given set of aspects,  the algorithm 
classifies the post as:  $\{\mbox{No Argument}, \mbox{Argument For}, \mbox{Argument Against}\}$. 
Tested on a publicly available benchmark, this algorithm exhibits superior performance: an F1 score of 0.6325 compared to 0.3796 of the earlier \texttt{bilstm} model \citep{stab-etal-2018-cross}.

\arman{All this is new} In an effort to evaluate the effectiveness of the method, we compared the BERT-based model against ChatGPT \cite{openai2023gpt4}. \miok{ChatGPT is arguably the most well-known   large language model currently, which has been proven to be able to understand complex and nuanced natural language with high accuracy.} Using the gold labeled validation data, (therefore unseen by the BERT-based model), the \citet{ukpclassification} argument classification model had an $F_{1}$ score of 0.6325. On the same data, ChatGPT had an $F_{1}$ score of 0.6330. Furthermore the Cohen's Kappa agreement score is 0.4567 indicating a moderate level of agreement between the two models. We can safely conclude that the argument classification model we use is performing at the current research and industry standard. In future works, we will focus on making improvements to this classification model.

\subsection{B. Our Argument Grouping Methods}

The goal here is to compare, cluster, and summarize arguments.
We specifically apply these methods to arguments in the discussion forums, but the methods can operate on any set of arguments.

    \subsubsection{Argument Similarity.}
    A fundamental capability is to identify posts that establish the same argument.
    To measure the similarity of arguments, we adopt and customize a BERT-based content similarity model which is shown to provide great results \citep{ukpclassification}. 
    The model is trained on 28 contemporary topics, from the UKP ASPECT Corpus using a fine-tuned BERT transformer. Upon evaluation, this method has been shown to achieve an F-score of 0.67 compared to 0.75 for humans.

\begin{algorithm}[tb]
\caption{Argument Clustering}
\label{alg:algorithm}
\textbf{Input}: A pair of arguments and their similarity score [0,1]\\
\textbf{Parameter}: None\\
\textbf{Output}: [0,n] amount of clusters containing similar arguments
\begin{algorithmic} 
\STATE Let $N=0$.
\STATE Let \textit{similarity\_threshold} = 0.75
\FOR{S1 and S2 in model output}
\IF {S1 \& S2 $>=$ \textit{similarity\_threshold}}
    \IF {S1 AND S2 not in existing clusters}
        \STATE Create new cluster and add S1 and S2
        \STATE $N$+=1
    \ELSIF {S1 in existing cluster but not S2}
        \STATE Add S2 to S1's cluster
    \ELSIF {S2 in existing cluster but not S1}
        \STATE Add S1 to S2's cluster
    \ENDIF
\ENDIF
\ENDFOR
\STATE \textbf{return} N 
\end{algorithmic}
\end{algorithm}

    \subsubsection{Argument Clustering.} We use the above similarity measure to cluster similar arguments.  For a group of arguments, we iterate through each argument, calculating the similarity to every other argument. We employ the following logic for the strict hierarchical clustering of arguments. We choose to derive strict non-overlapping clusters to get the true diversity of viewpoints.
    Once we have the similarity score for every pairwise combination of arguments, we  create cluster of  arguments which are similar. The algorithm for this is described in Algorithm~\ref{alg:algorithm}.

    \subsubsection{Summarizing  clusters of arguments.}
    Given our clustering, we wish to provide summaries of the substantive content of each cluster.  We use SBERT, a state-of-the-art modified BERT network using siamese and triple network structures that accurately computes the semantic meaning  of a sentence using sentence embeddings \citep{https://doi.org/10.48550/arxiv.1908.10084}. The evaluation of SBERT reveals significant improvement over state-of-the-art sentence embeddings such as GloVe embeddings and BERT out-of-the-box embeddings. Furthermore, SBERT is computationally efficient, performing tasks 55\% faster than Universal Sentence Encoder, a necessary requirement for corpuses as large as ours. Using a pre-trained sentence transformer, every sentence is converted to a 384 dimensional dense vector space before being passed through the SBERT summarization model. 

    \subsection{C. Our Thread Level Methods}

     Having identified arguments and their essence, we can model 
     the deliberation process at the level of a thread. In other words,
     we want to capture and quantify the structure of the \textit{dialogical} discussion
     and the interplay of its arguments.

\subsubsection{The deliberation profile of a thread.} We propose to profile a thread and its deliberation activity with the following metrics, some of which will be defined in this section:
a) number of posts, b) argumentative posts, c) depth of thread tree, d) sub-threads, e) \DSlong. 
Sub-threads is the number of times a post is responded to by more than one post.

    \subsubsection{Representing threads using graphs.}
    A key novelty of our work is the use of semantically-enhanced tree structures to represent the discussion that unfolds in a thread as shown in Figure~\ref{figs:thread_tree}.
    It should be clear that this representation: (a) has significant visual power,
    and (b) lends itself to quantifiable metrics leveraging graph theoretic concepts.

     We represent a thread as a directed graph, $   G = (V,A)$,
         where $V$ is a set of nodes which represent the posts, and $A$ is a set of directed edges. 
      Threads create a tree structure where each post $p_2$ is a response to some previous post $p_1$ in that thread, and  we use a directed edge, $(p_1, p_2)$ to connect the two posts. Only the first post that creates the thread does not have an outgoing edge. Furthermore, a node does not have to have an incoming edge, in other words a responding post.
    
    Let $v \in V$. We denote the in-degree of node $v$ by deg$^-(v)$ and the out-degree to be denoted as deg$^+(v)$. In our graph, it is possible to have nodes that do not have any incoming edges, or any outgoing edges, but can theoretically have as many as $A$ incoming 
    and zero or one outgoing edges (posts that start a thread have zero outgoing degrees).
    
    We use the term \textbf{Fan Out} to describe the number of leaves in a tree (thread). This captures the number of sub-threads or branches of discussion within a thread. For example, if  two posts independently reply to the first post in a thread the fan out value will be 2. The fan out of the thread in Figure~\ref{figs:thread_tree} is 7.
    For visualizing the threads, we use the open-source tool Gephi ~\citep{Bastian_Heymann_Jacomy_2009}.

    \subsubsection{Thread Deliberation Intensity.}
    We next propose a measure of the deliberative intensity of threads.  Intuitively, a thread has higher ``deliberative intensity" if it has: (a)  a greater variety of distinct arguments indicated by clusters of arguments, 
    and (b) a greater number of argumentative posts. 
    We introduce two metrics \Dargument and \Dcluster to capture each dimension of deliberation respectively.

    \begin{align}
 \textnormal{\Dcluster} &= \frac{\textnormal{\# Clusters}}{\textnormal{\# Arguments}} ,\begin{aligned} \textnormal{ for \# Arguments} > 0 \end{aligned}
       \\
        \textnormal{\Dargument} &= \frac{\textnormal{\# Arguments}}{\textnormal{\# Total Posts}},\begin{aligned} 
 \textnormal{ for \# Total Posts} > 0 \end{aligned} 
    \end{align}
    Intuitively, 
        cluster diversity captures the variation of arguments within a thread as the percentage of arguments that are unique.
     Similarly, argument diversity captures the argumentativeness of a thread as the percentage of posts that are argumentative.
     
    We now define our Deliberation Intensity Score, $\DS$, of a thread by considering both the above metrics  as follows:

\begin{equation}
        \DS =  \sigma_1 * \textnormal{\Dcluster}   +   \sigma_2 * \textnormal{\Dargument}
\end{equation}
Setting a value for the weights $\sigma_1$ and $\sigma_2$ allows us to put more emphasis on the diversity of interest. 
Here, we define their values with the use of logit functions:
$a_1 = \frac{1}{1 + e^{-(\textnormal{\# Arguments})}}$ and 
        $a_2 = \frac{1}{1 + e^{-(\textnormal{\# Total Posts})}}$.
We set $\sigma_1 = \frac{a_1}{a_1+a_2}$ and $\sigma_2 = \frac{a_2}{a_1+a_2}$.
The logit functions use the data to assign weights, with the intuition if the denominator of the ratio increases, the importance increases. For example, 30\% of unique arguments
out of 20 arguments is more ``important'' than the same ratio out of 5 arguments.

\subsubsection{Identifying important arguments within a thread.}
Having a graph representation, we can leverage powerful graph mining techniques to study the thread. 
One such application is identifying important arguments within a thread.
Adopting a graph theoretic approach, the importance
of a node in a graph reflects its connectivity: an important node will be connected to many other important nodes.
We propose, PostRank, a modified version of the PageRank algorithm to identify important nodes \citep{ilprints422}. The algorithm works by iteratively calculating the importance of the node by considering 
the importance of its neighbor nodes. 
Another consideration for PostRank is the bias PageRank will have towards the root node of trees, as this will always be the source of all incoming edges.   

To overcome the issue above, and give more on emphasis on the discourse, 
 we remove nodes with deg$^+(v) = 0$, before running PostRank. The node with the highest score is the post 
 that spins off more sub-threads, which is an indication of its impact and importance.

\subsubsection{Argument Stance Dependence.} We calculate the dependence of argument stances within a thread using the conditional probability: 
  \begin{equation}
      P(B | A) = \frac{P(B \cap A)}{P(A)}
  \end{equation} 
    \noindent Where $P(B | A)$ is the probability of the subsequent B argument stance, given the previous A argument stance. $P(B \cap A)$ is the probability both stances occur, while P(A) is the independent probability of argument stance A occurring. 

\section{Case Study: The GMO Debate on Reddit}

\begin{figure}[t] 
\centering
\begin{minipage}{\linewidth}
  \includegraphics[width=\linewidth]{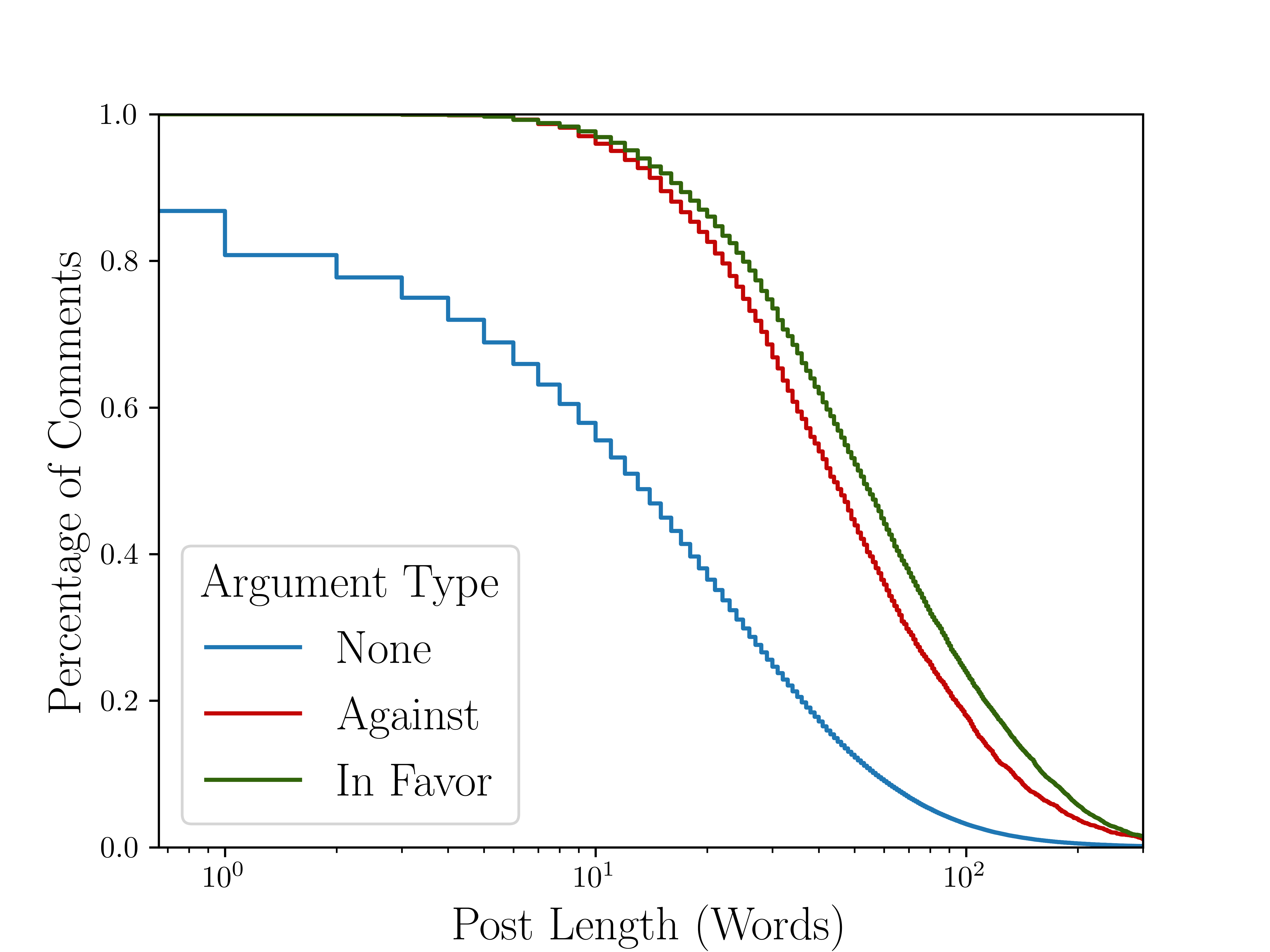}
  \centering
\end{minipage}
\caption{ The CCDF of the length of the posts in number of words for different types argument stance. Interestingly, posts in favor of an aspect tend to be longer than arguments against an aspect, while posts with no argument are the shortest.}  
\label{figs:postlengthcdf}
\end{figure}

We demonstrate the kind of results and insights we can get from our approach by applying it on the Reddit forums that we described earlier. 

To showcase the way our approach can be used in practice, we elaborate on the way we go about defining our topic and identifying its related aspects.
Recall that we are focusing on the GMO debate. We can define the topic as a set of keywords. Some of the aspects are given by experts, but this is optional. Our method can identify related aspects in an unsupervised way as we discussed earlier.

\begin{itemize}
    \item {\bf Topic}: GMO, Genetically Modified Organisms
    \item {\bf Aspects}: 
    
    {\bf a. Provided by experts:} gene editing, CRISPR, biotechnology, genomics based

    {\bf b. Identified by \algo:} Monsanto, China, John Deere, Climate Change, Soil Science

\end{itemize}

\subsection{Part 1: GMO Arguments and Argument Groups}

Here, we study properties of posts and the arguments that these posts may contain.

Our work is motivated by the intuition that online forums can
provide substantial information on deliberation around a topic of interest.
We first start by making two observations that support this thesis.

{\bf Argumentation happens:  \ArguPerc\% of posts contain arguments.} We find that across all the posts of the GMO threads \ArguPerc\% of them contain an argument. Arguments are essential for deliberation. At the same time, posts without an argument may still provide information,
as they can convey: (a) agreement or disagreement, or (b) provide information, e.g., pointing to a relevant article.

{\bf  Users put effort in their arguments: \PostLengthPerc\% of argumentative posts have more than \WordsPerPost words.} We observe that argumentative posts represent substantial effort. 
We use the length of
the post in number of words as a proxy for the amount of effort.
We plot the distribution of the post length for 
 for non-argumentative, In Favor argumentative, and Against argumentative posts in Figure~\ref{figs:postlengthcdf}.
We see that roughly \PostLengthPerc\% of argumentative posts have  
more than \WordsPerPost words.
Note that 50 words corresponds\footnote{https://numberofwords.com/character-count}
roughly to 250 characters. As reference a twitter post is currently limited to 280 characters, while its ``ideal" range  for user engagement is 71-100 characters (or roughly 20 words) according to marketing studies.
We argue that the higher number of words in the posts indicates
more effort, which possibly translates to more substantiated opinions.

\begin{table}[t]
\centering
\begin{tabular}{|l|c|c|}
\hline
    \textbf{Post Type} & \textbf{\#Posts} & \textbf{Percentage}\\ \hline
    
    All & 22,537  & 100\% \\ \hline 
     No Argument & 16,521  &  73\% \\ \hline
    With Arguments  & 6016 &  27\%\\ \hline \hline
    Arguments For & 3,796  &   17\% \\  \hline
    Arguments Against & 2,220  & 10\%\\ \hline
    
\end{tabular}
\caption{Quantifying argumentation: posts with arguments and further categorized as having a stance In Favor or Against towards the aspect that the argument is based on for our \ThreadsGMO dataset.}
\label{tab:arguments}
\end{table}

\subsubsection{Argument Upvotes.} 
We investigate the probability for the number of upvotes  each argumentative post receives, irrespective of their aspect. 
We discovered posts expressing an argument In Favor of an aspect have a higher likelihood of obtaining more upvotes than posts expressing arguments against an aspect.

Upvotes are an important metric to consider, as it is an indirect measure of how  accepted and trustworthy a comment is by users, although the pattern of upvotes might reflect a kind of bandwagon or herd behavior \citep{Ciampaglia2018-bw}. 
Furthermore, votes can magnify arguments 
yet also be tools for suppression as specific posts and arguments can be targeted, if they go against the community's beliefs  \citep{graham2021sociomateriality}. Therefore, we can conclude arguments of a positive or favorable stance may unintentionally be promoted by Reddit's ecosystem, while arguments being expressed with negative language are becoming less visible and accepted. This is an important distinction, as a post can be arguing against a topic while being in support of social good. For example, ``I am against chemicals that cause cancer.''   

{\bf Identifying important arguments.}
Recall that our PostRank algorithm can identify posts that seem to play a critical or central role in their discussion as captured by the thread.
Indicatively, we show results of this algorithm by listing some of the most influential posts.
This capability could be used to provide a fingerprint of the key posts within a thread and by extension of a forum. 

\begin{quote} ``Synthetic fertilizers and excessive tilling caused the dust bowl. I would look up no till gardening and stick with natural organic fertilizers. Even the guy who created synthetic fertilizers realized it was bad and refused to use it but the ease of use trumped all logic so it erupted into what is now modern gardening. Even my grandparents used to say organic is not better so I would not necessarily trust the old timer methods too much.'' 
\end{quote}
\noindent This was correctly classified as an argument against synthetic fertilizers. The next most important argument made was against GMOs: 
\begin{quote}
``The only concern I have with GMOs is if they are all the same potentially leading to a massive failure of crops should a disease or something hit a specific gene of the predominant strain. Kind of like if some disease suddenly wiped out all the holsteins the dairy industry would be devastated.''
\end{quote}

\begin{figure}[t] 
\centering
\begin{minipage}{\linewidth}
  \includegraphics[width=\linewidth]{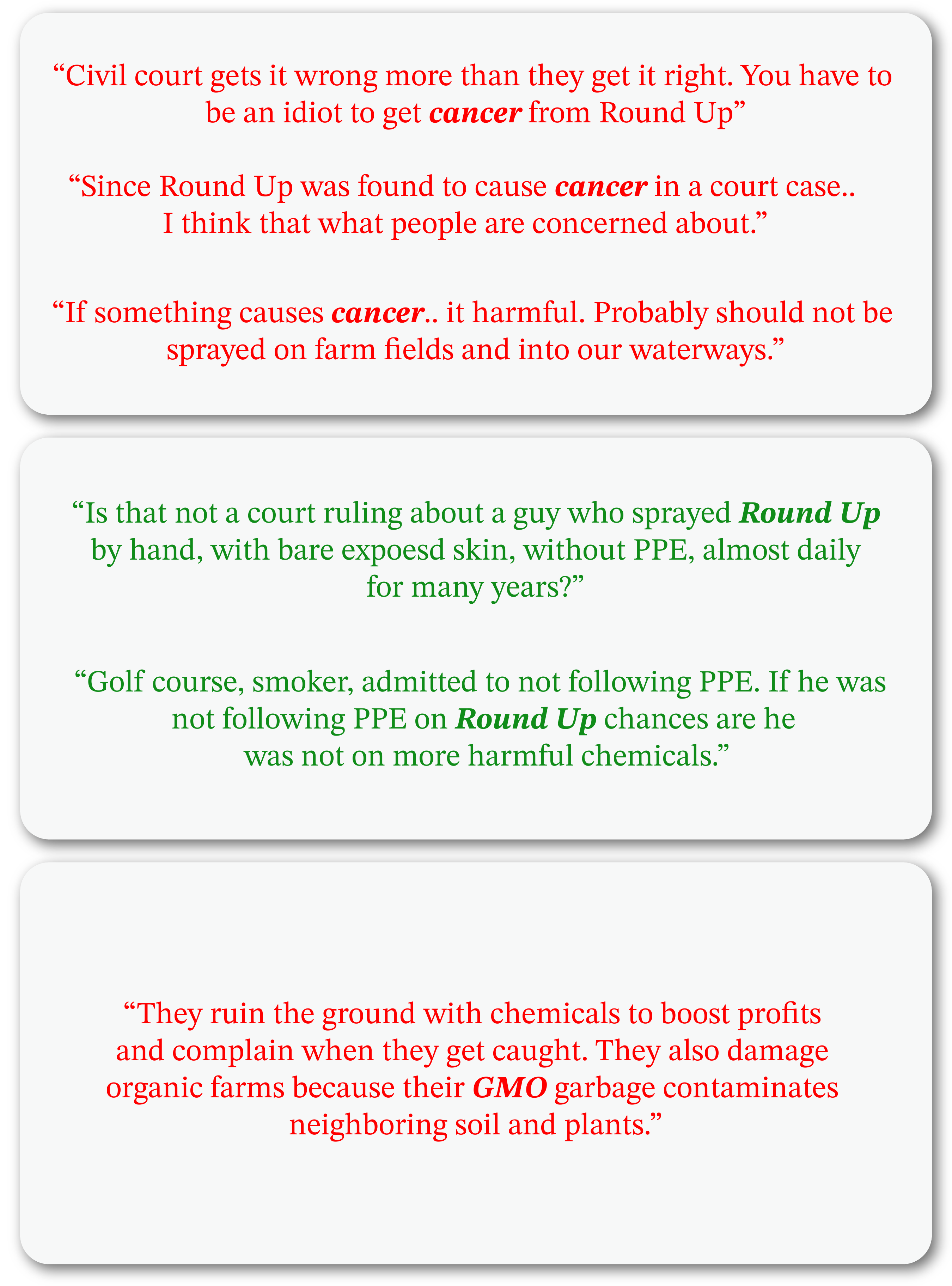}
  \centering
\end{minipage}
\caption{Using the arguments from the thread visualized in Figure \ref{figs:thread_tree}, we cluster semantically similar arguments together. The aspects are in bold italics, green colored texts represent arguments in favor of the aspect, and text in red represents arguments against the aspect.}  
\label{figs:Clusters}
\end{figure}

{\bf  Argument Clustering.}
 We use our argument clustering method as described in our Methodology section.
 Here we will use our \ThreadsXLong dataset with 74 threads, which just so happens to consist of 74 arguments.
 We run the argument similarity model across the threads. Then we apply our clustering method to group similar arguments. This reveals a relatively diverse set of arguments for GMOs, with 26 distinct clusters out of 74 total arguments If clustering based on topic and stance, In Favor of arguments for GMOs reveals 14 distinct clusters with 45 arguments while Against GMO arguments reveals 8 distinct clusters with 28 total arguments. Proportionally this reveals that In Favor arguments are more diverse than Against arguments.

{\bf  Argument Summarization.}
Once we have identified the most discussed arguments, and clustered similar arguments together how can we practically gain insight from what we have found? We propose to include the summarization of our clusters as an additional step in our framework's pipeline.

To illustrate this we display one of the summarized clusters for arguments In Favor of GMOs.
\begin{quote}
``If anything, GMO technology has potential to slow this trend down because it is much cheaper than traditional breeding which means it much easier for smaller players, like universities, to found new companies and increase competition innovation.''
\end{quote}
Conversely, summarizing of one of the clusters that contain against GMO arguments yields the following result:
\begin{quote}
``Over 80\% of genetically modified crops grown worldwide are engineered to tolerate being sprayed with glyphosate herbicides, 1 the best known being Roundup. Fuck Monsanto, and now Bayer, and i urge any idiot who thinks its harmless to get real comfortable with having nonhodgkins lymphoma...''
\end{quote}
The motive behind summarizing clusters is to quickly and effectively gain an understanding about what the general views and arguments are towards controversial topics. By doing so, policy makers and educators are able to identify misinformation or calls of concern. Given the two summarized clusters, we can gain a snapshot into the overarching narratives of the prevailing arguments.

\subsection{Part 2: Deliberation in GMO Threads}

\begin{figure}[t] 
\centering
\begin{minipage}{\linewidth}
  \includegraphics[width=\linewidth]{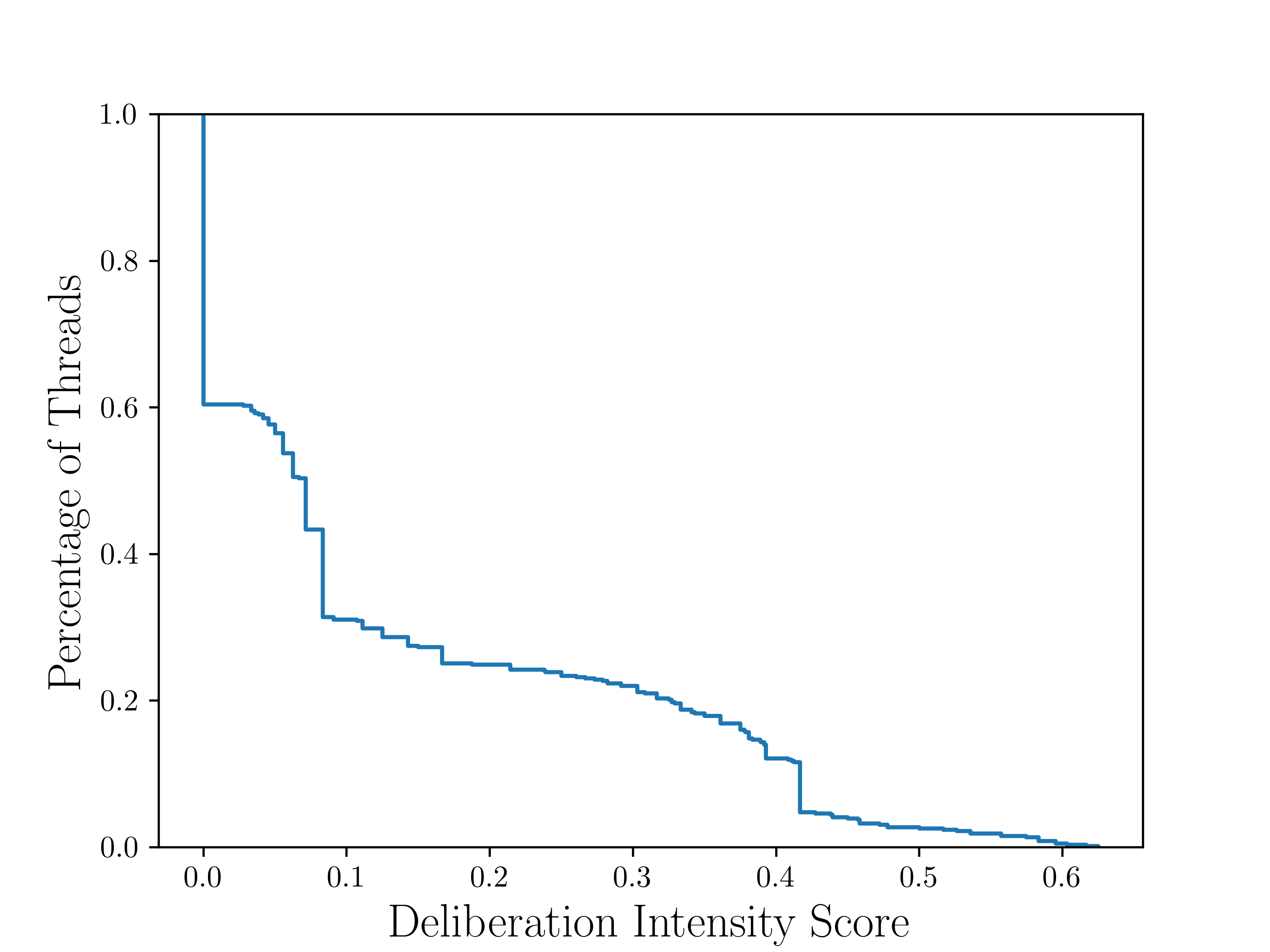}
  \centering
\end{minipage}
\caption{The CCDF of the \DSlong  distribution for \ThreadsLong (586 threads that have more than 5 posts.)}  
\label{figs:deliberation}
\end{figure}

\begin{table*}[t]
\centering
\begin{tabular}{|l|l|l|l|l|l|}
\hline
   \textbf{Dataset} & \textbf{\# of Nodes} & \textbf{\# of Edges} & \textbf{Avg. Degree} & \textbf{Avg. Tree Depth} & \textbf{Max Tree Depth}\\ \hline
      \ThreadsGMO  & 22,537 & 13,960 & 0.680 & 1.6 & 12 \\ \hline 
\end{tabular}
\caption{Properties of the graph representations of the threads in our \ThreadsGMO dataset. }
\label{tab:TGMO-dataset}
\end{table*}

Here, we apply our methods to modeling the dialogical dynamics of the deliberation in our online forum. We want to quantify the deliberativeness of the threads
and understand the interplay of arguments in the discussion.

In our \ThreadsGMO dataset, we have  8,316 threads with 22,537 posts.
Additional properties of this dataset can be seen in Table~\ref{tab:TGMO-dataset} and recall that we represent a thread as a tree.
For example, we see that the average tree depth is 1.6 while the maximum tree depth is 12.

We want to see the distribution of argumentative posts across the threads. In Figure~\ref{figs:histogramposts} we can see the breakdown of the argument stances across all of our aspects. Soil Science is the aspect with the higher number of arguments, with more arguments In Favor of the aspect compared to arguments Against.

{\bf Modeling the post-response dynamic.} 
\miok{A natural question that arises is whether a correlation exists between arguments and their stance between consecutive posts in online dialogue. This is a large topic in its own right and here we will explore only one dimension. Specifically, we focus on the users' stance towards the aspect in their post. This analysis focuses on \textit{per-post} basis, which may not capture full  \textit{conversational level} nuances. 
}

\miii{Let's create a smoother transition to the next paragraph: which "this conditional probability"? Let's start. We start by calculating/measuring xxxx and xxxx. We then calculate the probability that xxxx... OR if thsi si already define earlier. "Specifically, we use the probability/xxx as defined in XXX.}

The formulation for this conditional probability is outlined in our Methodology section. Applied on our dataset, we observe that if the post is an argument against an aspect, then there is 59\% likelihood of the response also being an argument against the same aspect compared to 40\% probability of being In Favor. For posts with arguments In Favor of an aspect, the response stance is equally likely to be in In Favor and Against, the aspect 50\% and 49\% respectively.

We can conclude from this that Arguments In Favor of an aspect tend to result in more stance diverse arguments, while Arguments Against an aspect tend to create a bias towards more against arguments being generated as responses.

\begin{figure}[t] 
\centering
\begin{minipage}{\linewidth}
  \includegraphics[width=\linewidth]{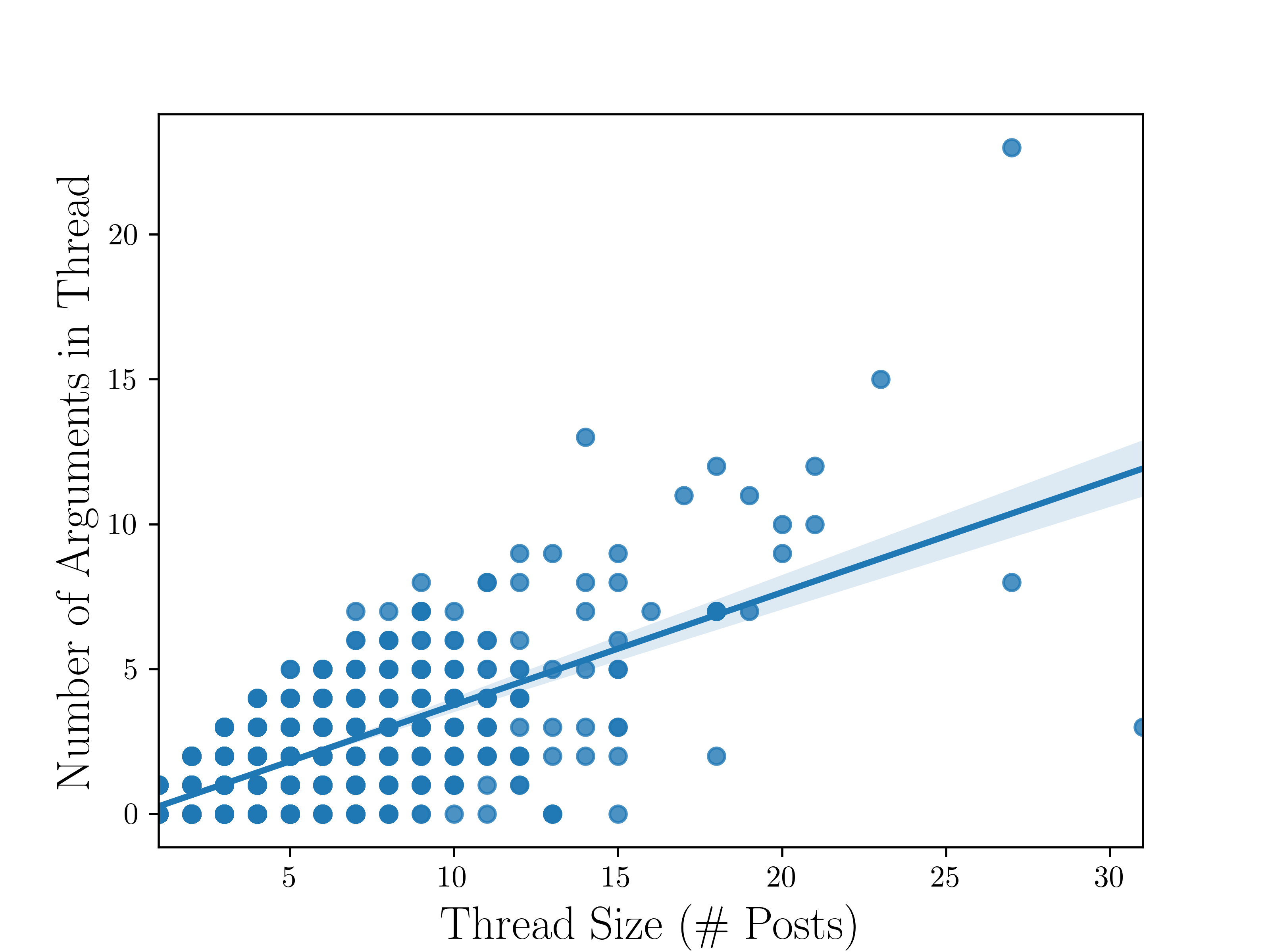}
  \centering
\end{minipage}
\caption{Number of argumentative posts versus the size of the thread and showing also the line of a Linear Regression fit.}
\label{figs:thread_length}
\end{figure}

{\bf  Argumentation and Deliberation.}
We want to understand which threads are likely to generate more deliberation. We plot the number of argumentative posts as a function of the size of the thread in Figure~\ref{figs:thread_length}.
We observe that there is a loose correlation between the two metrics as indicated by the linear regression fit with 95\% confidence.
This is somewhat expected as more posts are more likely to have more arguments.

\begin{figure}[t] 
\centering
\begin{minipage}{\linewidth}
  \includegraphics[scale=0.52]{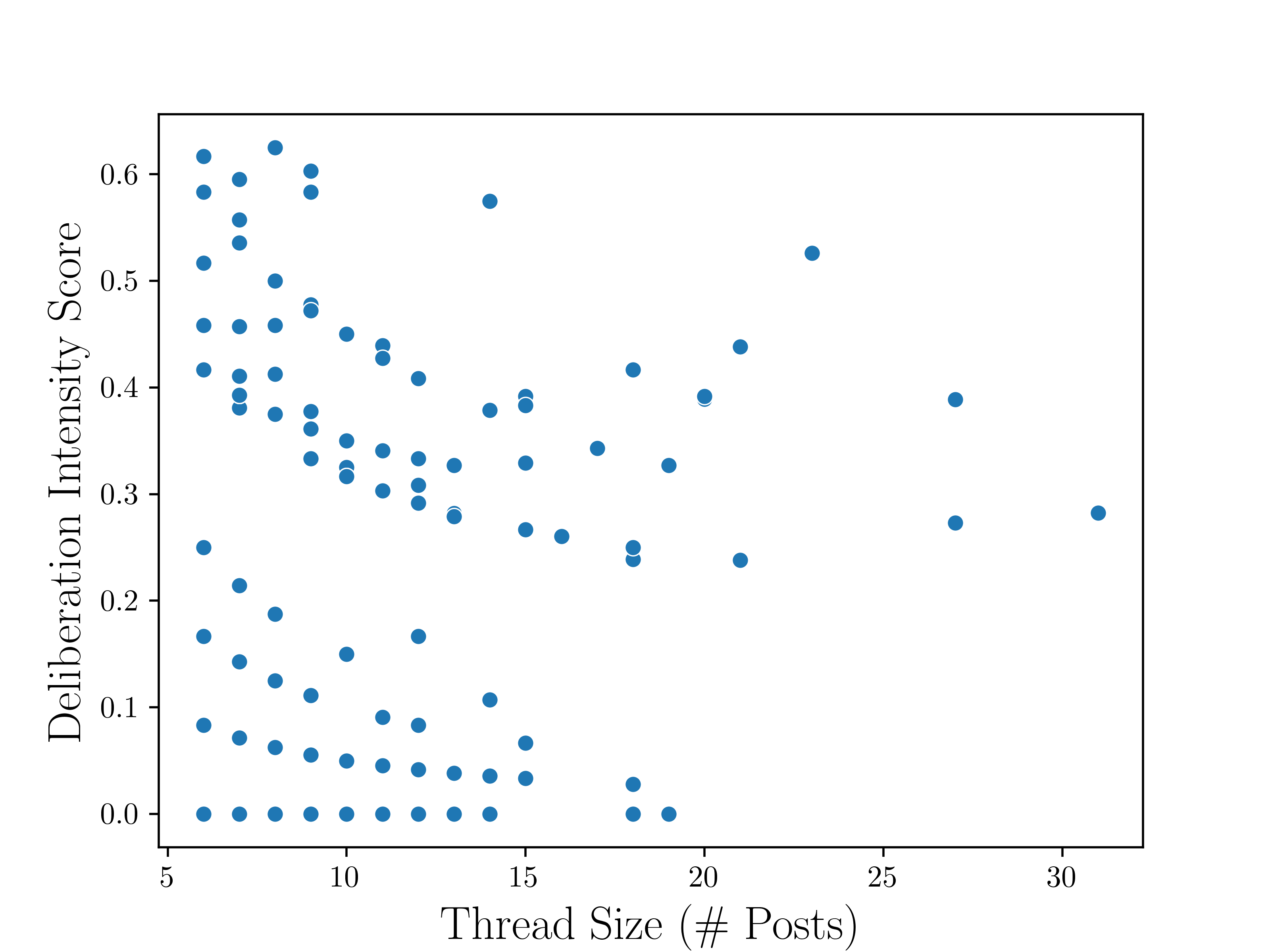}
  \centering
\end{minipage}
\caption{The \DSlong as a function of the size of a thread. There is no clear correlation between thread size and deliberation intensity, which
is rather counter-intuitive. \arman{New:} Our initial manual inspection of large threads and low Deliberation Intensity \miok{ attributes this phenomenon to the presence of }  Question \& Answer type discourse. Future work will investigate these discourse nuances more comprehensively. }  
\label{figs:deliberation_intensity}
\end{figure}

{\bf \DSlong is not affected by thread size.}
We arrive at this counter-intuitive observation by 
assessing the quality of the deliberation
with our \DSlong. Recall that this metric considers the percentage of argumentative posts and thus removes the advantage of longer threads compared to shorter threads.
We plot the
\DSlong of threads versus their size in Figure \ref{figs:deliberation_intensity}: the two metrics are not correlated. This indicates that there is a hidden variable that our deliberation intensity score is capturing, 
and that is the relative intensity of the argumentation (by considering the percentage of argumentative posts), and the diversity of these arguments (using the number of argument clusters).

\section{Discussion: Perspectives and Ethics}

{\bf Broader perspective: understanding versus polarization.}
We see our approach as a significant capability in mining social media.
The goal of our work is to understand discourse: the interaction between users and their arguments at the level of a discussion.
As such, our work leverages and incorporates existing methods, and its value is that it synthesizes to create a holistic framework for understanding discourse.
We envision our work as an enabling capability that can: (a) quantify online deliberation, and (b) reveal not only the opinions that users hold toward policy topics, but also the reasoning for why they hold those opinions \citep{Mercier2012-ui}.
Its ideal application is to help promote dialogue and understanding as a way to bridge differences in controversial topics \citep{Sunstein2017-st}.
Reducing polarization requires listening to and understanding the diverse reasons and arguments supporting different perspectives \citep{Cohen1989}.
This helps to ensure all diverse views in society have a voice, and we can move towards constructive communication.

Having said this, we should stress that technology is ultimately a tool, which can be used for good or evil: a hammer can be used to build a house or injure someone. Malicious users can find ways to exploit our framework, including the ability to identify important arguments, to raise the visibility of morally objectionable content in order to sow discord or spread their viewpoint.

\miok{ {\bf Applicability and limitations.} 
The applicability of our framework is fairly wide.
It can be used for: (a) a wide range of topics, and (b) most online discussion forums.
Naturally, specific features of a forum or a topic could limit the type of information that can be extracted.
First, our framework could apply to many topics, especially ones that can be anchored by a set of keywords.
Given one or more keywords that describe a topic of discussion,
such as ``abortion,'' ``police brutality,'' ``war on Ukraine,'' we can use these, or a union or intersection of these, to identify aspects and arguments around the topic.
In addition, we can use well established keyword-expansion methods to identify the appropriate keyword sets~\cite{IKEA-Joobin-2022}.
Second, our framework can apply equally well to all discussion forums or even commenting platforms.
Obviously, the usefulness of our thread-level analysis relies on the features that we saw in Reddit, which include forums forming communities of interest, with the post as a well-defined discrete argument unit, and the opportunity to observe users offering posts in response to other posts. Our thread-level measures would only apply to other online forums with similar affordances. Luckily, these features are widely supported by most discussion platforms.
  By contrast, some platforms differ in substantial ways. For example, users are anonymous in some forums like 4chan, which will prohibit a user-centric analysis of the argumentation.
}

\miok{Going beyond online forums, \algo can be used for different sources of data, such as  transcripts of dialogue. In fact, we intend to study congressional hearings and compare the argumentation differences between politicians and lay people in online forums.
}

{\bf Public dissemination and open-sourcing.} 
We intend to share our code and datasets with the research community. We will provide a GitHub repository that will provide the code, a how-to manual, and select datasets, including our labeled data, which can greatly facilitate future efforts. We also have a funded partnership with the news aggregator website \texttt{AllSides.com} to provide results from our validated methods for the public on a variety of policy topics.

\textit{Beyond English.}The current implementation of our framework works on English text. However, it is easy to see how it can extend to other languages. Adapting to a new language requires changes in the text processing capabilities. Specifically, the adaptation is quite straightforward for alphabet-based languages, as the key thing is to train the models with the new language. Note that there are BERT models for logographic and other types of languages such as Chinese\cite{Cui_2021}, and Arabic \cite{antoun2020arabert}.

\miok{
{\bf Accuracy and biases.} The effectiveness of our framework relies on several factors.
First, the quality of the data is a critical factor, as poorly written posts and badly formed statements can become impediments.
Second, our approach relies on the effectiveness of the methods that we deploy in each step of our pipeline. The positive view  here is that as better methods emerge, we can incorporate them in our framework and benefit from them.
Third, the keyword selection, which we consider as an input, can introduce biases or limit the representativeness of a study.
}

{\bf Ethical Considerations.}
Our work follows ethical guidelines as specified by the AAAI code of conduct and ethics. 
First, the data that we collected is publicly available: anyone can access the discussions on Reddit and users know that their comments are visible by anyone who visits the platform. Furthermore, IRBs do not require informed consent for data collection \cite{vitak2017ethics}.
Second, all the data is anonymized, and any metadata 
linking to personal identifiable information
was removed in the preprocessing stage. 
Finally, all reported data is aggregated in a way that there are no users identified by name.

\section{Related Work}
\label{sec:related}

While there is extensive work developing methods for unsupervised topic and argument mining, most of this work centers on classifying arguments in an individual’s \textit{monological} speech or writing.  By contrast, our work leverages monological classifiers to measure the inter-user \textit{dialogical process} within online forums. Our work differs from the majority of previous efforts by virtue of its focus that rests at the intersection of: (a) online forums, (b) argument extraction, and (c) quantifying and assessing the dialogical process of argumentation within and between forum threads.

For recent reviews of argument mining, see \citet{Lawrence2020-na} and \citet{Schaefer2021-yt}. There is a vast literature in computational linguistics and computer science centered on mining of monological arguments from a variety of corpuses, including structured text such as legal writing \citep{Palau2009-ja}, news articles \citep{Ein-Dor2020-hd, Slonim2021-hp}, and student essays \citep{Stab2017-da}, as well as unstructured text mostly from web-scraped or Wikipedia data \citep{ukpclassification, Chernodub2019-cc, Daxenberger2020-zx, Habernal2017-ve, Stab2018-qg}. 

Within the field of argument and stance detection there are several works utilizing NLP to classify different niche grammatical devices. \citet{liu2022politics} uses pretrained language models to characterize and predict ideologies across different genres of text. \citet{zhang2022kcd} uses graph neural networks to classify political perspectives. \citet{conforti2022incorporating} uses multi-genre stance detection for financial signaling. \citet{sia2022offer} uses a constraint based modeling approach to predict the winning argument in Reddit debate datasets. 

There are relatively few efforts however, to leverage these monological classifiers to measure and assess the dialogical process of argumentative discourse in online forums like Reddit using unsupervised methods. At the dialogical level, much of the work measuring the nature and quality of argumentation in social science and communication uses either hand coding or semi-supervised methods to assess transcripts of speeches \citep[such as][]{Fournier-Tombs2020-ca}. Most similar to our work is \citet{Chakrabarty2020-eq} who identify intra-turn argument relations between posts on Reddit, which we complement by proposing macro-level measures of the nature and quality of the dialogical argumentative process within and between threads. We note other papers making use of argument extraction graph representation \citep[for example][]{Sakai_undated-xm} and that leverage topic and stance detection and other tools from opinion mining \citep{Stede2020-vh}. Dialogical methods for assessing the \textit{persuasiveness} of arguments \citep[such as][]{Boschi2021-km, Dutta2020-hr} are complementary to our work but outside of the scope of this paper.

\section{Conclusion}

The key contribution of our work is \algo, a comprehensive and systematic approach for modeling deliberation in online forums. 
The approach consists of many capabilities that rely on specific model and metric definitions.
Specifically, we develop methods: (a) detecting argumentative posts; (b) describing the structure of arguments within threads with powerful visualizations; and (c) summarizing and clustering arguments,
and (d) quantifying the deliberation diversity of threads. 

We conduct a substantial study, where we apply our approach on four communities on the Reddit platform over a span of \duration months
which produced more than 200K posts and more than 35K users.
We focus on the GMO debate in order to have a long lasting and
reasonably controversial topic.
 We find that \ArguPerc\% of posts contain arguments and these posts are often fairly long.
 
 Our observations suggest that there is substantial dialogue that happens in these forums. We qualify this, although we note that we selected threads that relate to a controversial topic. Nevertheless, the case study demonstrates the opportunity our methods present to mine dialogical argumentation as researchers and institutions an gain insights into not only the opinions people hold, but also how
 people reason about a topic in terms of both arguments and how they debate their position.

{\em Future Work.} Encouraged by this initial study, we intend to extend our work in two different directions. First, we will improve, expand and fine-tune our methodology. Second, we will expand our study to: (a) more topics, and (b) more online discussion forums.

\section{Acknowledgments}
This work was supported by NSF SaTC Grant No. 2132642.
and CDFA grant 2021 Specialty Crop Block Grant Program H.R. 133  No. 000705282.

\vspace{.2em}

\bibliography{mybib}

\end{document}